\documentclass[english,11pt]{article}
\usepackage[T1]{fontenc}
\usepackage[latin9]{inputenc}
\usepackage{graphicx}
\usepackage{blois,epsfig}
\usepackage{subfig}
\usepackage{babel}

\begin{document}
\vspace*{4cm}

\title{QUBIC, a bolometric interferometer to measure the B modes of the
CMB}

\author{Jean Kaplan}
\address{APC, Université Paris-Diderot-Paris 7, CNRS\\
on behalf of the QUBIC collaboration \footnote{France: APC Paris, CESR Toulouse, CSNSM Orsay, IAS Orsay; India: IUCAA Pune; Ireland: NUI Maynooth; Italy: Università di Milano - Biccoca, Università di Roma - la Sapienza, Università di Sienna; UK: University of Manchester; USA: Brown University Providence, University of Wisconsin  Madison, Richmond University, Richmond; URL: http://qubic.org.}}

\maketitle

\abstracts{Measuring the B modes of the CMB polarization fluctuations would provide
very strong constraints on inflation. The main challenge in this measurement
is the treatment of systematic effects. CMB observations with imagers
and interferometers, subject to very different systematics, are complementary
in this respect. Interferometry provides direct access to the Fourier
transform of the sky signal. In bolometric interferometry,  the interference pattern produced by the sky through a few hundred horns is imaged on a bolometer array. Several such modules are needed to achieve the required sensitivity.
We will describe QUBIC, a merger of the US and European MBI and BRAIN
collaborations.  QUBIC is a polarized bolometric interferometer to be deployed
in 2011-2012.}

\section{CMB polarization}
The CMB polarization results from the local quadrupole in the distribution
of last scattered photons. One can separate the polarization patterns
in the sky according to their parity: even parity $E$ modes, and
odd parity $B$ modes. For reasons of symmetry, scalar primordial
fluctuations only produce $E$ modes, whereas tensor perturbations
(gravity waves) produce both $E$ and $B$ modes, therefore detection
of $B$ modes means detection of gravity waves.
\paragraph*{Inflation}
In most inflationary models the dominant scalar fluctuations are adiabatic
and the level of $E$ modes can be inferred from the level of temperature
fluctuations, as has been observed in the WMAP observations \cite{WMAP5:PowSpecabr}.
In contrast, the Tensor/Scalar ratio $r=T/S$ is not known but
depends on the rate of production of primordial gravitons which is
related to the energy scale of inflation by $r\simeq0.1\left(\frac{E_{\mathrm{inflation}}}{2\times 10^{16}\mathrm{GeV}}\right)^{4}$.
At the moment, $r$ is only bounded above by the temperature power
spectrum : $r<0.33$ \cite{Brown:2009abr}. The best upper bound from
upper limits on polarization $B$ modes is weaker \cite{Chiang:2009abr}.
\section{\textit{B }mode detection}
$B$ modes are at least ten times weaker than $E$ modes but could
be much smaller. Therefore, to measure them we need i) very high
sensitivity, ii) extremely good control of systematics and iii)
an efficient and reliable foreground subtraction. To make sure that
$B$ modes have really been detected, several converging observations
with different systematics and different foreground subtraction methods
would be beneficial.\\

\paragraph*{Two techniques} have so far been used to measure the CMB polarization
fluctuations:\\
{\em Imagers} : this technique has been the most commonly used. It provides excellent sensitivity, in particular if one uses bolometers which
allow to reach the photon noise limit. The forthcoming bolometer arrays
will allow to detect very low level CMB polarization fluctuations.
However it suffers from several kinds of systematics: i) Telescope
induced systematics (beam, polarization angles, ground pick up); ii)
polarization measurements requires taking differences between detectors,
which is a source of intercalibration errors; iii) as the sky image
is obtained from its time ordered scanning by a narrow beam, time
variations of the atmosphere emissivity and transparency are critical.\\
{\em Heterodyne interferometry} is the first technique to have detected
CMB polarization \cite{kovac02:dasi}. In interferometry, i) no front  optics
is necessary; ii) no differences between detectors are needed to measure
polarizations; iii) interferometry directly observes the Fourier transform
of the sky within the beam, and therefore is not sensitive to time
variations of the atmosphere. However, in heterodyne interferometry
the noise is dominated by the amplification noise, which does not
allow reaching the photon noise limit. Moreover it seems technically
difficult to measure the numerous baselines needed to reach the sensitivity
required for $B$ mode detection.\\
{\em Bolometric interferometry} may combine the advantages of both
techniques

\section{Bolometric interferometry}

In bolometric interferometry, bolometers do not see the sky directly
but they see the interference pattern produced by the sky through
an array of horn antennas. The signal collected by each horn is separated into
its two orthogonal polarization components, one of the polarization
is rotated by $90^{\circ}$ so that they can interfere, and each channel
is appropriately phase shifted. The signals
from all different channels are combined in a beam combiner to produce
the interference pattern. The combined signals illuminate a bolometer
array. Bolometric interferometry is an additive interferometry which
means that the visibility, the Fourier transform of the sky signal,
must be separated from the total bolometer output. This can be done
by modulating the phase shifts.

The difficulty is to separate the $2N(N-1)$ complex visibilities
corresponding to all pairs of the $N$ input horns and 2 polarizations.
When $N$ is large, an electronic solution to this problem is desperately
complex. This is the reason why we turned to an optical solution.

\paragraph*{The Quasi Optical Interferometer (QOI)}

The principle of the QOI is the following (see figure \ref{fig:QOI})
\cite{Timbie:2006}: the signal from each polarization and each horn
is re-emitted by 2$N$ back horns, and the interference pattern of
the back horns is collected by a bolometer array on the focal plane
of an internal telescope. Thus, all parallel rays re-emitted by all
back horns in the same direction reach the bolometer array on the
same pixel. The additional geometrical phase shifts create the interference
pattern. 

As has been shown in two dedicated papers \cite{Char:2008,hyland09}, the real and imaginary
part of all visibilities can be separated in an optimal way using time sequences
of phase shift modulations.

\paragraph*{Sensitivity}

Compared to an imager with the same number of input channels, the
sensitivity of a bolometric interferometer is roughly worse by a factor
2, as can be seen on figure \ref{fig:Sensitivity-a}. This is the
price to pay for better or in any case different systematics. A detailed
comparison has been published \cite{Hamilton08a}.

\section{QUBIC}

The US and European collaborations MBI and Brain have decided to join
their efforts and build a large interferometer, QUBIC, aimed at measuring the
$B$ modes of the CMB. A possible configuration would be as follows: 6 modules, each equipped with 144 input horns, ($\simeq10000$ baselines)
and 288 back horns in a compact square array and a focal plane of
$\simeq$ 900 Transition Edge Sensors. The primary beams would have a FWHM of $\sim 14^{\circ}.$ To check for foreground contimination
we would have 3 frequency channels : 90, 150 and 220 GHz, with 25\%
bandwidth. The cryogenics would involve a 4K pulse cooler for each module
and a 100-300 mK dilution unit for the focal plane. The multipole
range would be approximately $25\leq\ell\leq150$. Such an instrument
would allow to reach $r\sim0.01$ in one year of continuous operation
at DOME C in Antarctica. The expected QUBIC sensitivity compared to some
other ongoing or planed experiments is displayed in figure \ref{fig:sensitivity-b}

\paragraph*{Present status and tentative schedule:}

Before merging into QUBIC in 2009, the Brain and MBI collaboration worked independently.\\
\textbf{The Brain collaboration}  has launched two site testing
(atmosphere, logistics) path finder campaigns at Dome C during the
antarctic summers of 2006 and 2007, and a winter-over campaign is
scheduled starting during the antarctic summer 2009\\
\textbf{The MBI collaboration} has built a four horn prototype interferometer,
MBI-4, and taken data in 2007 and 2008. Fringes have have been observed
with MBI-4.\\
\textbf{The QUBIC collaboration} intends to deploy a first module
of QUBIC, possibly at DOME C, in 2011/2012. The goal is to complete the full QUBIC instrument by 2014 at the latest.

\begin{flushleft}
\begin{figure}[p]
\begin{centering}
\includegraphics[scale=0.4]{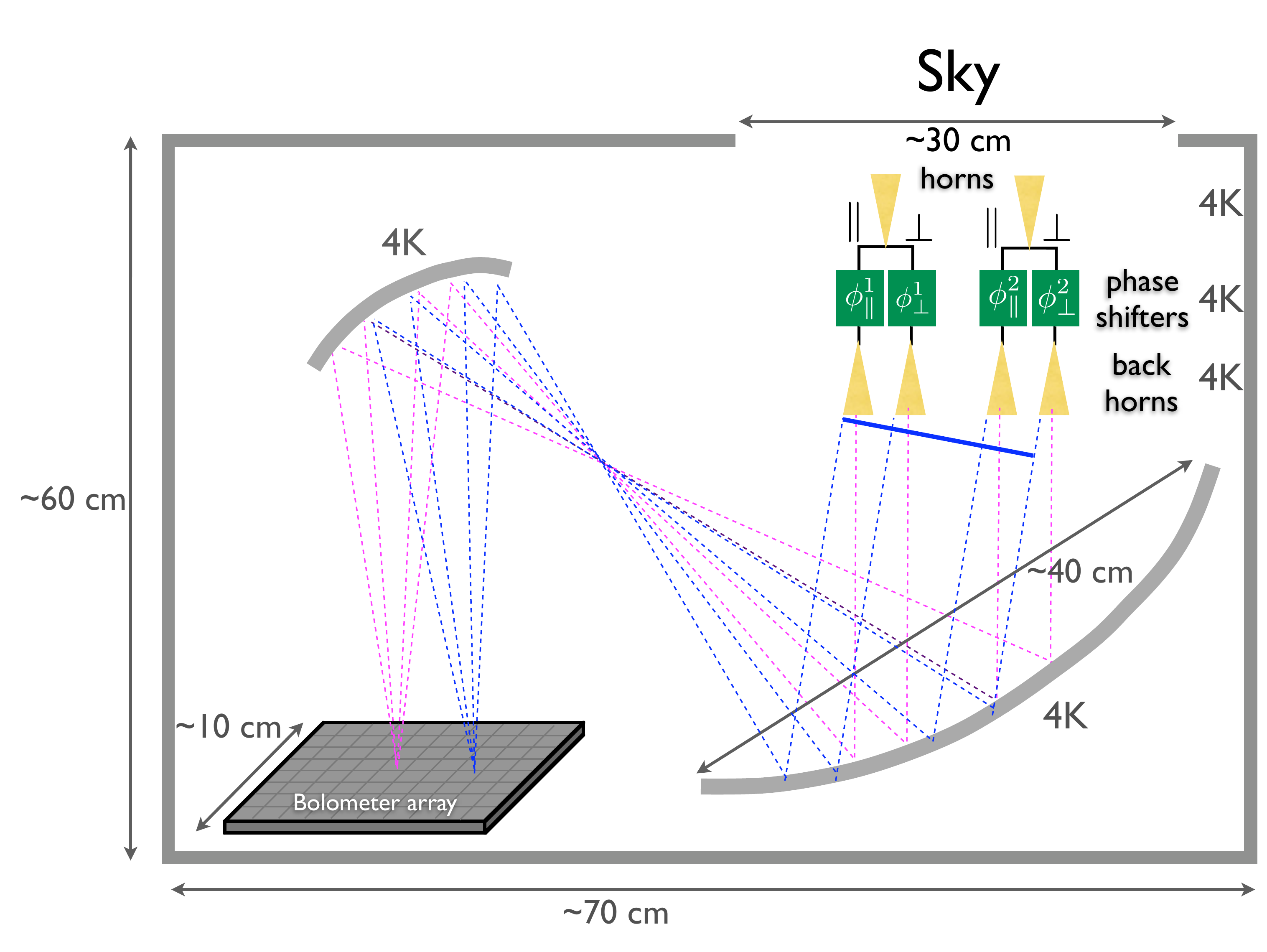}
\par\end{centering}
\caption{The Quasi Optical Interferometer\label{fig:QOI}}
\end{figure}

\begin{figure}[p]
\subfloat[Sensitivity of a bolometric interferometer compared with an imager
and an heterodyne interferometer. Thick curves : noise only; thin curves include cosmic variance with $r\sim 0.1$ \protect\cite{Hamilton08a}. \label{fig:Sensitivity-a}]{
\includegraphics[scale=0.33]{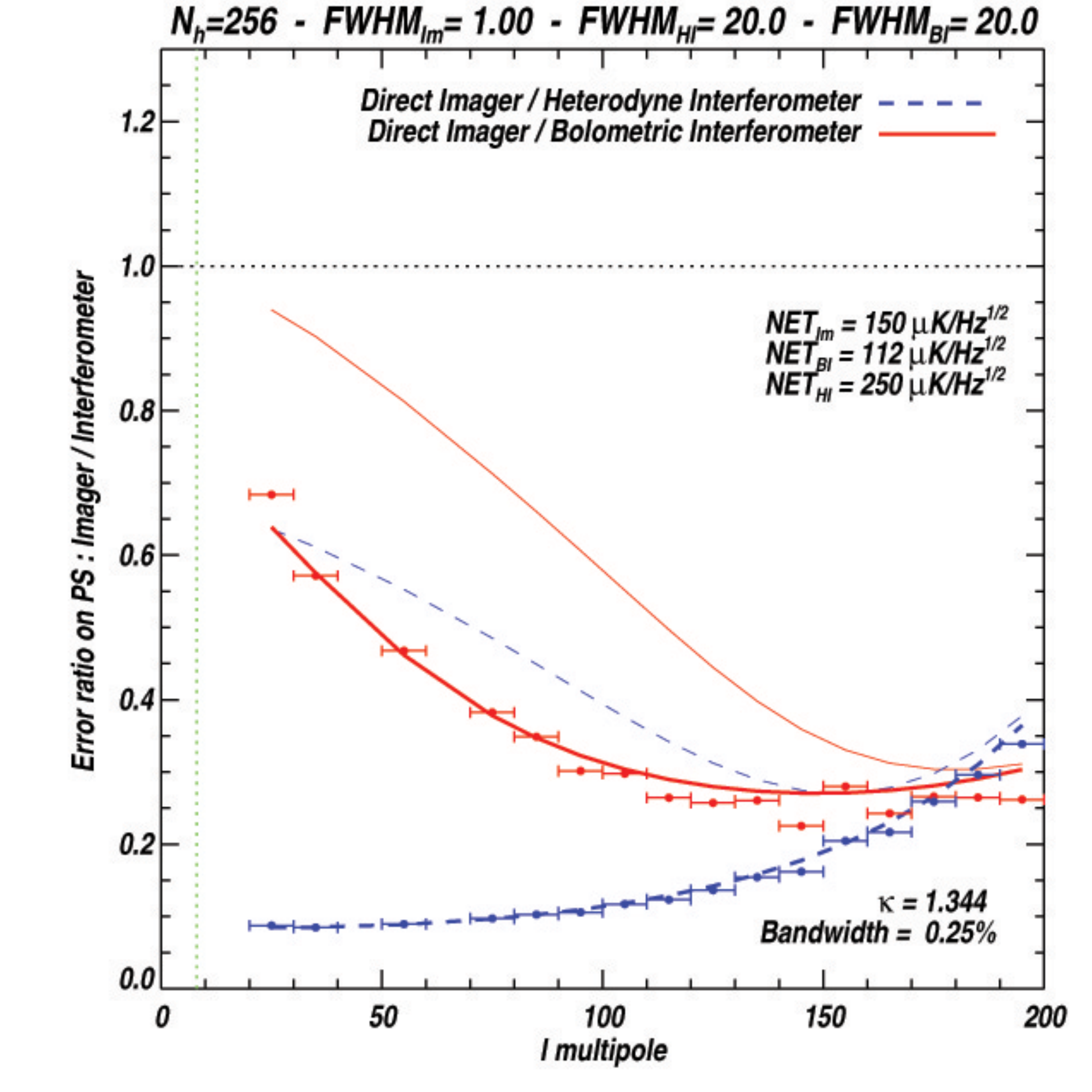}
}\subfloat[QUBIC sensitivity compared with Ebex, Clover and Planck \protect\cite{Hamilton08a}. \label{fig:sensitivity-b}]{\begin{centering}
\includegraphics[scale=0.17]{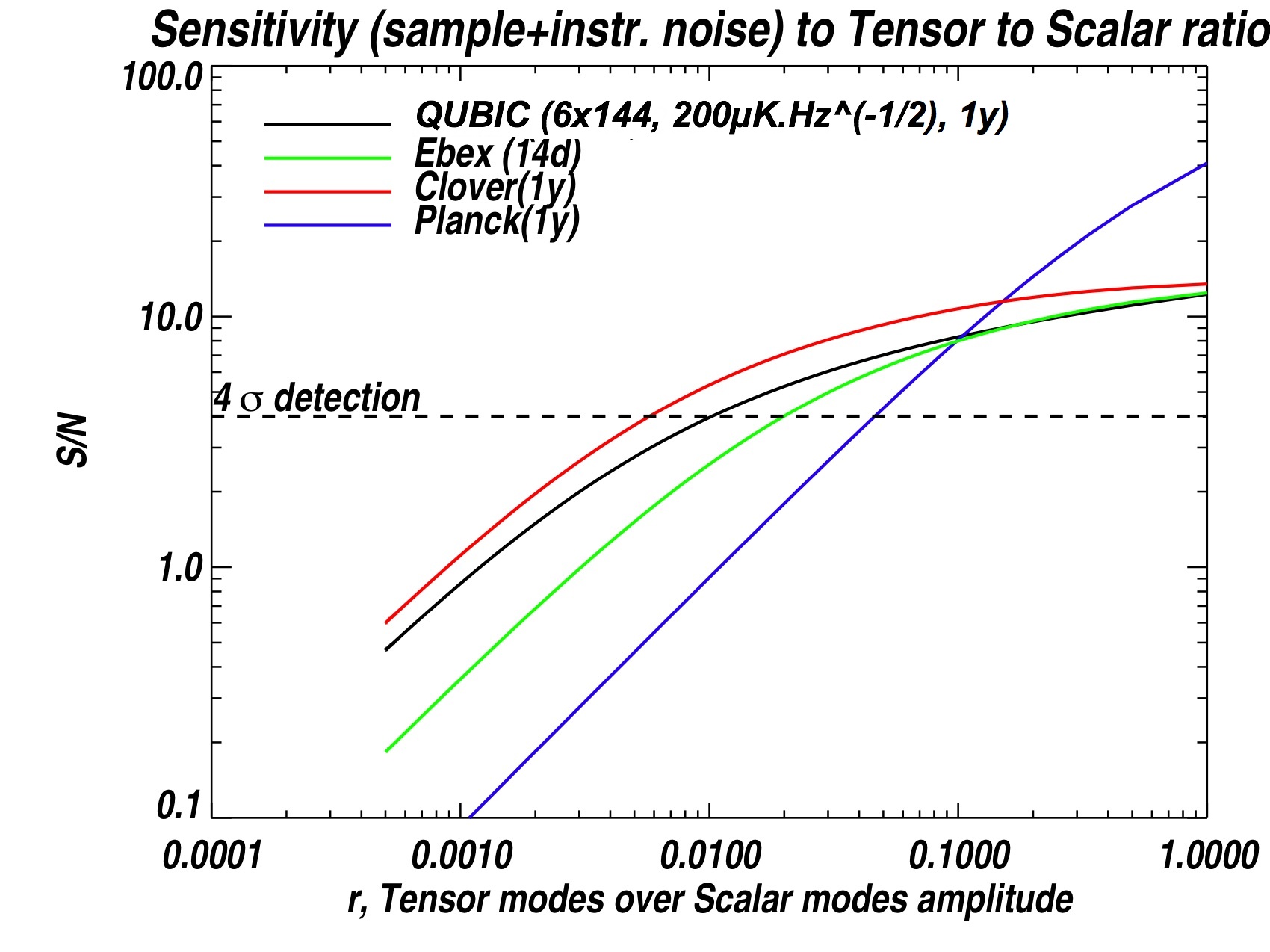}
\par\end{centering}
}\caption{} 
\end{figure}

\section*{References}
\bibliographystyle{unsrt}
\bibliography{/Users/kaplan/Library/texmf/bibtex/bib/brain,/Users/kaplan/Library/texmf/bibtex/bib/cmb,/Users/kaplan/Library/texmf/bibtex/bib/CMBPAPER}

\par\end{flushleft}
\end{document}